\documentclass[aps,prd,twocolumn,superscriptaddress,secnumarabic]{revtex4-2}

\usepackage[total={7.2in,9in}]{geometry}
\usepackage{amsmath, amssymb, amsfonts} % mathematical formulae
\usepackage{graphicx}  % insert images
\usepackage[colorlinks, linkcolor=blue, citecolor=blue, urlcolor=blue]{hyperref}
\usepackage{bm}
\usepackage{physics}
\usepackage[utf8]{inputenc}
\usepackage{geometry}
\usepackage{amsmath}
\usepackage{fancyhdr}
\usepackage{graphicx}
\usepackage{framed}
\usepackage{amsthm, amssymb, appendix, bm, graphicx, mathrsfs}
\usepackage{orcidlink}

\newcommand{\jhu}{William H.\ Miller III Department of Physics and Astronomy, Johns Hopkins University, \\ 3400 N. Charles Street, Baltimore, Maryland, 21218, USA}
\newcommand{\ipmu}{Kavli IPMU (WPI), UTIAS, The University of Tokyo, Kashiwa, Chiba 277-8583, Japan}

\begin{document}

\title{Off-Equatorial Orbits around Magnetically Charged Black Holes}

% author information
\author{Xilai Li\orcidlink{0009-0004-4125-1057}}
\email{martianian@sjtu.edu.cn}
\affiliation{Zhiyuan College, Shanghai Jiao Tong University, Shanghai, 201210, People's Republic of China}

\author{Loris Del Grosso\orcidlink{0000-0002-6722-4629}}
\email{ldelgro1@jh.edu}
\affiliation{\jhu}

\author{David E. Kaplan\orcidlink{0000-0001-8175-4506}}
\email{david.kaplan@jhu.ed}
\affiliation{\jhu}
\affiliation{\ipmu}

\date{\today}

\begin{abstract}
We present a complete characterization of stable, off-equatorial circular orbits around magnetically charged black holes (MBHs). For a static, spherically symmetric MBH, we derive an exact analytic expression for the orbital latitude $\theta$ as a function of radius $r$ and we %establish a direct connection between these orbits and the spacetime fundamental structure, and 
analyze the effect of synchrotron radiation. We show that charged particles such as electrons and protons can exhibit $\mathcal{O}(1)$ latitude deviations at the innermost stable circular orbit (ISCO) radius and remain stable under synchrotron emission even for extremely small values of the black hole magnetic charge. We then extend the analysis to rotating MBHs, numerically computing the prograde and retrograde orbital branches and demonstrating how frame-dragging modifies their structure and stability regions. We show that these off-equatorial orbits are a unique feature of the magnetic charge, being forbidden in the analogous electrically charged Kerr-Newman spacetime.
Our results suggest that environments surrounding magnetically charged black holes can exhibit distinctive phenomenological signatures, potentially offering a way to constrain the magnetic charge of astrophysical black holes.
\end{abstract}

\maketitle % generate title

% Section 1: introduction
\section{Introduction}
The coupling of magnetic charge to gravity provides a rich theoretical laboratory for probing classical and semiclassical black-hole physics \cite{Lee1992, MBH2,Bai:2020spd,Bai:2020ezy,Gervalle:2024yxj}.  Magnetic monopole arise naturally in many extensions of the Standard Model, particularly in grand-unified theories (GUTs), where they are predicted to form as topological defects during symmetry-breaking phase transitions in the early universe \cite{Dirac1931, tHooft1974, Polyakov1974}. Their gravitational counterparts, magnetically charged black holes (MBHs), serve as an idealized framework for studying how non-trivial gauge fields alter spacetime geometry and the dynamics within it, including the causal structure, particle motion, and radiative processes \cite{Gibbons1975, Lee1992, Tursunov2018, Rayimbaev2021}. A key feature of these spacetimes is the non-zero divergence of the magnetic field, which prevents the electromagnetic vector potential from being defined globally. Nevertheless, a local potential can be defined in regions free of the monopole source, allowing for a consistent analysis of charged particle dynamics \cite{Rohrlich1966, Preskill1984}. Furthermore, MBHs are connected to their electrically charged cousins—the Reissner–Nordström and Kerr–Newman black holes—via electromagnetic duality, a fundamental symmetry of Maxwell's equations where the roles of electric and magnetic fields can be interchanged \cite{duality1, duality2}. Studying the phenomenology of MBHs is therefore of intrinsic theoretical interest and may also inform searches for observational signatures of such compact objects~\cite{Liu:2020vsy, Liu:2020bag, Chen:2022qvg,Grilli:2024fds} or distinguish them from exotic mimickers~\cite{Bah:2025vbr}.

The analysis of particle and light trajectories around black holes is a cornerstone of relativistic astrophysics \cite{orbit1, orbit2, orbit3, orbit4}. The paths of light are described by null geodesics, which determine the properties of gravitational lensing and define the location of the photon sphere, a region composed of unstable circular light orbits that is fundamental to the appearance of a black hole's shadow \cite{Synge1966, Bardeen1973, Chael2021}. The paths of massive objects are described by timelike geodesics, which govern the dynamics of accretion disks and define the innermost stable circular orbit (ISCO). The ISCO represents the minimum radius at which a massive particle can stably orbit a black hole before beginning an inevitable plunge, making it a critical parameter in models of accretion efficiency and compact-object inspirals \cite{Bardeen1972, CompactInspiral, ISCO1, ISCO2}. The dynamics become significantly richer when the test particles are electrically charged. The Lorentz force, acting in concert with gravity, can produce qualitatively novel phenomena, such as strong latitudinal deflection and the emergence of stable, off-equatorial orbits not possible for neutral particles~\cite{Carter1968, Pugliese2011, Grunau2011, Hackmann2013,Russo:2020lah,Kasuya1982, Pereniguez:2024fkn}. Such off-equatorial motion has a classical, non-relativistic precedent: Poincaré showed that a charged particle in the field of a magnetic monopole follows a conical trajectory \cite{Poincare1896}, and the quantum mechanics of such systems was later developed in detail \cite{Zwanziger1968}. A comprehensive analytic characterization of circular bound orbits around static and rotating MBHs, together with an analysis of their stability under electromagnetic emission, 
%and a connection between their properties and the fundamental structure of the spacetime, 
is still lacking in the literature. The present work aims to fill this gap.

In this paper, we analyze the orbital structure of neutral and charged test particles in two key backgrounds, adopting natural units where $c=G=1/4\pi\epsilon_0=1$. The first is the static, spherically symmetric MBH, described by the metric function
\begin{equation}
f(r)=1-\frac{2M}{r}+\frac{P^2}{r^2},
\label{eq:static_metric_f}
\end{equation}
where $M$ is the mass and $P$ is the magnetic charge. The second is its rotating counterpart, a Kerr-Newman MBH parameterized by $M$, $P$, and specific angular momentum $a=J/M$ \cite{Reissner1916, Kerr1963, Bardeen1970, Adamo2014}.
In Section~\ref{sec:static}, we introduce the static, spherically symmetric MBH and derive a closed-form expression for the polar angle $\theta$ of charged particles on stable circular orbits.
Section~\ref{sec:isco}  links features of the $\theta$–$r$ relation to the photon ring and ISCO, and analyzes the effect of synchronton radiation on the stability of the orbits.
In Section~\ref{sec:rotate} we generalize to rotating MBHs and determine the prograde and retrograde orbital branches numerically.
In Section~\ref{sec:conclusion} we conclude.  We have included an appendix with the full equations of motion for a charged particle in a circular orbit around a rotating MBH (ignoring synchrotron radiation).

% Section 2: hypotheses
\section{Orbital Structure in Static MBHs} \label{sec:static}

We begin with the static, spherically symmetric MBH, whose geometry is described by the line element in Schwarzschild-like coordinates $(t,r,\theta,\phi)$:
\begin{equation}
    ds^2 = -f(r)\,dt^2 + f(r)^{-1} dr^2 + r^2 d\Omega^2,
    \label{eq:static_metric}
\end{equation}
where the metric function $f(r)$ is given by Eq.~\eqref{eq:static_metric_f} and $d\Omega^2 = d\theta^2 + \sin^2\theta \, d\phi^2$ is the line element on the unit 2-sphere. This metric is the magnetic analogue of the Reissner–Nordström metric and constitutes an exact solution to the Einstein–Maxwell equations for a purely radial magnetic field $\mathbf{B} = (P/r^2)\,\hat{\mathbf{r}}$ \cite{Reissner1916}. The cosmic censorship hypothesis imposes the bound $|P| \leq M$ \cite{Penrose1965, Krolak1986}, for which the spacetime possesses an event horizon ($r_+$) and an inner Cauchy horizon ($r_-$) at
\begin{equation}
    r_{\pm} = M \pm \sqrt{M^2 - P^2}.
\end{equation}
In the extremal limit, $|P| \to M$, these horizons coincide at $r_\pm = M$.

Due to the magnetic charge, the electromagnetic four-potential $A_{\mu}$ cannot be globally defined. It can, however, be expressed locally on gauge patches. For the northern hemisphere ($0 \leq \theta < \pi$), we adopt the standard form \cite{Rohrlich1966, Preskill1984}:
\begin{equation}\label{eq:non-rotating_potential}
    A^{(N)}_{\phi} = P(1 - \cos\theta), \qquad A^{(N)}_{t} = A^{(N)}_{r} = A^{(N)}_{\theta} = 0.
\end{equation}
The potential on the southern hemisphere, $A^{(S)}$, is related by a gauge transformation, $A^{(N)} - A^{(S)} = 2P \, d\phi$. Henceforth, we work in the northern hemisphere and omit the superscript $(N)$. The only non-vanishing components of the covariant electromagnetic tensor $F_{\mu\nu} = \partial_\mu A_\nu - \partial_\nu A_\mu$ are $F_{\theta\phi} = -F_{\phi\theta} = P\sin\theta$.

The spacetime possesses two Killing vectors: a timelike Killing vector $k^\mu = (1,0,0,0)$ associated with time-translation invariance, and an azimuthal Killing vector $\xi^\mu = (0,0,0,1)$ associated with rotational symmetry. For neutral particles, these symmetries guarantee the conservation of energy $E$ and angular momentum $J$.

For photons ($m=0, q=0$), which satisfy the null condition $g_{\mu\nu}u^\mu u^\nu = 0$, circular orbits define the photon sphere. Its radius, $r_{\text{ps}}$, is found by extremizing the effective potential, yielding
\begin{equation}
    r_{\text{ps}} = \frac{3M \pm \sqrt{9M^2 - 8P^2}}{2}.
    \label{eq:static_photon_sphere}
\end{equation}
where $r_{\text{ps}}^{+}$ and $r_{\text{ps}}^{-}$ correspond to the outer and inner photon spheres, respectively. The outer sphere represents the physically relevant photon sphere, and its radius decreases monotonically from the Schwarzschild value of $3M$ to $2M$ as $|P|$ increases from $0$ to $M$.

For massive neutral particles ($m \neq 0, q=0$), the innermost stable circular orbit (ISCO) marks the boundary between stable and plunging trajectories. The ISCO condition in the magnetically charged spacetime parallels that of the Reissner–Nordström case. The effective potential is given by
\begin{equation}
    V_\text{eff} = f(r)\left(1+\frac{j^2}{r^2}\right),
    \label{eq:static_Veff}
\end{equation}
where $j$ is the angular momentum per unit mass of the particle. The ISCO is determined by the standard marginal stability conditions,
\begin{equation}
    \frac{d V_\text{eff}}{dr} = \frac{d^2 V_\text{eff}}{dr^2} = 0.
    \label{eq:ISCO_condition}
\end{equation}
which, after solving for $j^2$, yields the cubic equation
\begin{equation}
    M r^3 - 6 M^2 r^2 + 9 M P^2 r - 4 P^4 = 0.
    \label{eq:static_ISCO}
\end{equation}
The real root of Eq.~\eqref{eq:static_ISCO} gives the ISCO radius $r_{\text{ISCO}}$, which decreases monotonically from the Schwarzschild value $r_{\text{ISCO}} = 6M$ to $4M$ in the extremal limit.

\begin{figure}[htbp]
  \centering
  \includegraphics[width=0.8\linewidth]{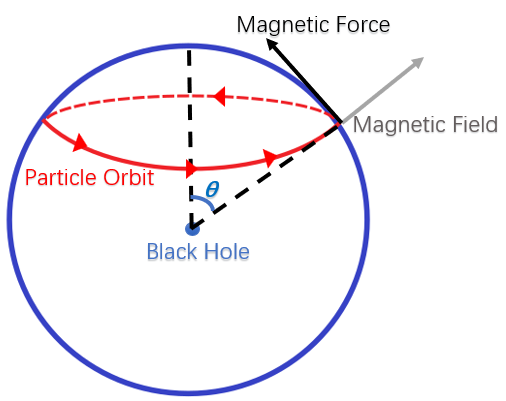} % File must exist!
  \caption{Circular orbit of a charged particle around an MBH. The magnetic force shifts the particle to a nontrivial polar angle.}
  \label{fig:Orbit_illustration}
\end{figure}

For a test particle with rest mass $m$ and charge $q$, its 4-velocity
\begin{equation}
    u^{\mu} = (\dot{t}, \dot{r}, \dot{\theta}, \dot{\varphi}),
\end{equation}
satisfies the constraint
\begin{equation}
    u^{\mu}u_{\mu}=\frac{\dot{r}^2}{f(r)}+r^2\dot{\theta}^2 +r^2\sin^2(\theta)\dot{\varphi}^2-f(r)\dot{t}^2 =-1\label{eq:normalize_velocity_general}. \\  
\end{equation}
Where the over-dot denotes a derivative with respect to proper time $\tau$.
The equations of motion can be derived from the Lagrangian
\begin{equation}
    \mathcal{L}=-\frac{m}{2}u^{\mu}u_{\mu}+qu^{\mu}A_{\mu}=-\frac{m}{2}u^{\mu}u_{\mu}+qA_\varphi\dot{\varphi},
    \label{eq:static_charged_lagrangian}
\end{equation}
which leads to
\begin{equation}
    \frac{d^2x^{\mu}}{d\tau^2} + \Gamma^{\mu}_{\alpha\beta} u^\alpha u^\beta = \frac{q}{m} F^{\mu\nu} u_{\nu},
\label{eq:static_eq_motion}
\end{equation}
where $\Gamma^{\mu}_{\alpha\beta}$ are the Christoffel symbols.

The presence of electric charge of the test particle introduces the Lorentz force that alters the orbital plane, displacing the trajectory from the equator. Despite this, stable circular orbits still exist for certain polar angles, as shown in Figure~\ref{fig:Orbit_illustration}.

For the test particle to move on a circular orbit, its 4-
velocity simplifies to
\begin{equation}
    u^{\mu} = (\dot{t}, 0, 0, \dot{\varphi}),
\end{equation}
and $f(r)\dot{t}^2 - r^2\dot{\varphi}^2=1$.
The $t$ and $\varphi$ components of Eq.~\eqref{eq:static_eq_motion} vanish identically as
\begin{equation}
    \begin{split}
&\Gamma^{t}_{tt}=\Gamma^{t}_{t\varphi}=\Gamma^{t}_{\varphi\varphi}=0, \\
&\Gamma^{\varphi}_{tt}=\Gamma^{\varphi}_{t\varphi}=\Gamma^{\varphi}_{\varphi\varphi}=0, \\
&F^{tt}=F^{t \varphi}=F^{\varphi t}=F^{\varphi \varphi}=0.
\label{eq:static_vanishing_quantities}
\end{split}
\end{equation}

The $r$ and $\theta$ components then determine the polar angle and angular velocity for a given radius. We begin with the $r$-component, which has no Lorentz force term. Using the relevant Christoffel symbols

\begin{equation}
\begin{split} 
 &\Gamma^r_{t\varphi}=0 ,\\  
 &\Gamma^r_{tt} = (1-\frac{2M}{r}+\frac{P^2}{r^2})(\frac{M}{r^2}-\frac{P^2}{r^3}),\\
 &\Gamma^r_{\varphi\varphi} = -r\sin^2\theta(1-\frac{2M}{r}+\frac{P^2}{r^2}),
 \label{eq:static_chr_r}
\end{split}
\end{equation}
we obtain the angular velocity
\begin{equation}
u^\varphi=\omega_{\pm}=\pm\frac{1}{r \sin\theta}\sqrt{\frac{\frac{M}{r}-\frac{P^2}{r^2}}{1-\frac{3M}{r}+\frac{2P^2}{r^2}}}.
\label{eq:static_angular_velocity}
\end{equation}
where $\omega_+$ and $\omega_-$ correspond to prograde (meaning counterclockwise as viewed from above the northern
hemisphere) and retrograde motion (opposite), respectively.

For the $\theta$ equation, the electromagnetic contribution yields
\[
\frac{q}{m}F^{\theta\nu}u_{\nu} = \frac{q}{m}g^{\theta\alpha}g^{\nu\beta}F_{\alpha\beta}u_{\nu}=\frac{q}{m}\frac{P}{r^2}\omega\sin\theta,
\]
while the geodesic term is
\[
\Gamma^{\theta}_{\alpha\beta}\frac{dx^\alpha}{d\tau}\frac{dx^\beta}{d\tau} = \Gamma^{\theta}_{\varphi\varphi}u^{\varphi}u^{\varphi}=-\omega^2\sin\theta\cos\theta.
\]
Combining these, the only possible circular orbit for the charged particle lies at 
\begin{equation}
    \tan(\theta_{\pm})=\mp\frac{mr}{qP}\sqrt{\frac{\frac{M}{r}-\frac{P^2}{r^2}}{1-\frac{3M}{r}+\frac{2P^2}{r^2}}}.
\label{eq:static_theta_r}
\end{equation}

These orbits exist only when the quantity under the square root is positive, which requires the radius to be larger than the photon-ring radius, as will be shown in Section~\ref{sec:isco}. For a charged particle located at $r_{\text{ISCO}}$, 
Eq.~\eqref{eq:static_theta_r} is accurately approximated by the expression
\begin{equation}\label{eq:scalingtheta}
\tan(\theta_{\pm})=\mp\,\, 2\sqrt{3} \,\,\bigg(\frac{q}{m}\bigg)^{\!\!-1}
    \!\!\left(\frac{P}{M}\right)^{\!\!-1} \,.
\end{equation}
This expression indicates that a realistic particle, such as an electron with a charge-to-mass ratio of $\sim 10^{22}$, can still have $\tan(\theta) \sim \mathcal{O}(1)$ for values of $P / M$ as small as $10^{-22}$, highlighting the strong influence that even a tiny magnetic charge can exert on light charged particle. This is somehow reminiscent of the findings in Ref.~\cite{Pereniguez:2024fkn}, where the authors show that even a single magnetic monopole in the black hole can significantly affect the superradiant instability timescale.

\begin{figure}[htbp]
    \centering
  \includegraphics[width=\linewidth]{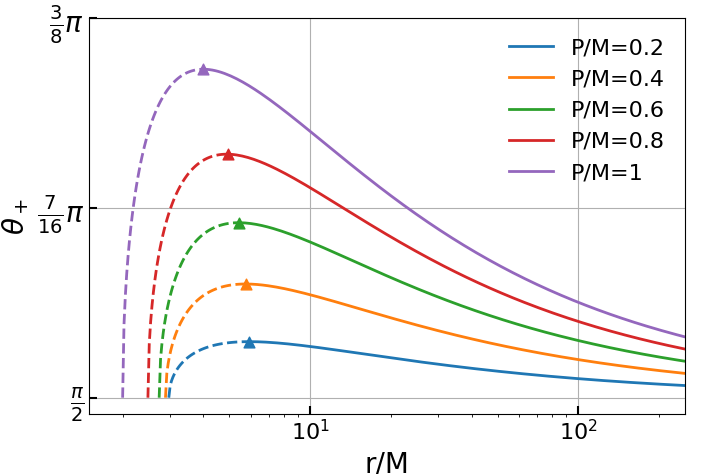} % File must exist!
  \caption{$\theta_+(r)$ of a $q/m=-1$ particle, for multiple $P/M$ MBHs. Solid lines denote stable orbits, while dashed lines indicate instability against small perturbations; the transition point (ISCO) is marked on each curve. Note that $\theta_+<\pi/2$ since the magnetic force on a prograde, negatively charged particle points above the equator. The y-axis is inverted so that north pole direction ($\theta = 0$) appears at the top of the figure, matching the intuitive orientation.}
  \label{fig:static_theta_r}
\end{figure}

Figure~\ref{fig:static_theta_r} shows $\theta_+(r)$ for a $q/m=-1$ particle and various magnetic charge-to-mass ratios of the MBH. Here $\theta_+<\pi/2$ because the magnetic force on a prograde, negatively charged particle is directed above the equator, decreasing its polar angle.  
From Eq.~\eqref{eq:static_theta_r}, such orbits exist only for $r$ larger than a certain threshold. The angle approaches $\pi/2$ in both the small- and large-$r$ limits, corresponding to equatorial motion. There is always a single peak in $\theta(r)$, whose location will be shown in Section~\ref{sec:isco} to coincide with the ISCO radius, while the smallest orbital radius matches the photon-sphere radius. The stability of these orbits against radiative decay will also be considered in Section~\ref{sec:isco}.

Although $\theta \to \pi/2$ as $r \to \infty$, the particle does not necessarily approach the equatorial plane. The vertical distance from the plane, $y = r|\cos\theta|$, can be found from Eq.~\eqref{eq:static_theta_r}:
\begin{equation}
    y = r|\cos\theta|=\frac{r}{\sqrt{\tan^2\theta + 1}}.
\label{eq:static_y}
\end{equation}
In the large-$r$ limit, the asymptotic behavior of $y$ is
\begin{equation}
    y \sim \frac{r}{\sqrt{\frac{m^2r^2}{q^2P^2}\frac{M}{r} + 1}} \sim \sqrt{r} \to \infty.
\label{eq:static_y_infinity}
\end{equation}
Thus, the orbit remains at an asymptotically diverging vertical distance from the equatorial plane, despite that $\theta \to \pi/2$.  

\begin{figure}[htbp]
  \centering
  \includegraphics[width=\linewidth]{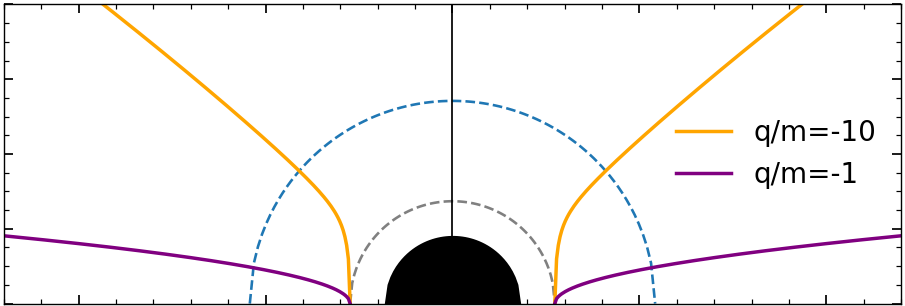}
  \caption{Cross-sectional view of circular orbits around a magnetically charged black hole with $P/M=0.6$. The black disk marks the event horizon, while the grey and blue dashed circles denote the photon sphere and the ISCO of neutral particles, respectively. The orange and purple curves indicate a series of orbits of prograde particles with $q/m=-1$ and $q/m=-10$.}
  \label{fig:static_show_orbit}
\end{figure}

Figure~\ref{fig:static_show_orbit} shows the upper half of a cross section of a MBH with $P/M=0.6$, together with representative prograde circular orbits for test particles with $q/m=-1$ and $q/m=-10$. The curves in the $(r,\theta)$ plane indicate a continuous series of such orbits. The black disk marks the horizon, while the grey and blue dashed circles denote the photon sphere and the ISCO radius, respectively.

% Section 4: Evidence for neutrino 
\section{Analysis of Orbits in Non-Rotating MBHs} \label{sec:isco}

We now perform a detailed analysis of the off-equatorial circular orbits derived in the previous section and assess their stability.

\subsection{Radial Boundaries and Stability Analysis}\label{subsec:static_isco}

The existence of the charged-particle orbits described by Eq.~\eqref{eq:static_theta_r} depends on the reality of the square-root term. For radii outside the event horizon ($r > r_+$), the numerator ($M/r - P^2/r^2$) is positive. The denominator, however, vanishes at
\[
    r = \frac{3M+\sqrt{9M^2-8P^2}}{2},
\]
which is precisely the photon-sphere radius given in Eq.~\eqref{eq:static_photon_sphere}. For a stable orbit to exist, the denominator must be positive, which requires $r > r_{\text{ps}}$. Thus, the photon sphere marks the innermost boundary for these off-equatorial circular orbits. As an orbit approaches this boundary from above ($r \to r_{\text{ps}}^+$), Eq.~\eqref{eq:static_theta_r} shows that $\tan(\theta) \to \infty$, implying $\theta \to \pi/2$. This has a clear physical interpretation: at the photon sphere, the particle's velocity approaches the speed of light, and its trajectory becomes equatorial, mimicking a null geodesic.

Next, we investigate the innermost stable circular orbit (ISCO) of a charged particle.
The time independence of the Lagrangian (Eq.~\eqref{eq:static_charged_lagrangian}) guarantees conservation of energy:
\begin{equation}
    E_\text{particle}=-\frac{\partial\mathcal{L}}{\partial u^t} = -mu_t=mf(r)\dot{t}
    \label{eq:static_charged_full_energy}. \\  
\end{equation}
Working per unit rest mass, we define
\begin{equation}
    \varepsilon=\frac{E_\text{particle}}{m} = mu_t=f(r)\dot{t}
    \label{eq:static_charged_energy}, \\  
\end{equation}
so that the temporal term in Eq.~\eqref{eq:normalize_velocity_general} becomes $-\varepsilon^2/f(r)$.

The angular terms require more care, as neither $u^\theta$ nor $u^\varphi$ is separately conserved. However, the combination representing the total angular momentum magnitude per unit mass, is found to be conserved. To prove it, we define
\begin{equation}
j=u^{\theta}u_{\theta} + u^{\varphi}u_{\varphi}=r^2[\dot{\theta}^2 +\sin^2(\theta)\dot{\varphi}^2].
\end{equation}
The total derivative of $j$ is:
\begin{equation}
    \frac{dj}{d\tau}=\dot{g}_{\theta\theta}(u^{\theta})^2 + 2g_{\theta\theta}u^{\theta}\dot{u}^{\theta} + \dot{g}_{\varphi\varphi}(u^{\varphi})^2+2g_{\varphi\varphi}u^{\varphi}\dot{u}^{\varphi}.
    \label{eq:absolute_angular_momentum_derivative} 
\end{equation}
where we have
\begin{equation}
    \begin{split}
    &\dot{g}_{\theta\theta}=\frac{d(r^2)}{d\tau}=2r\dot{r},\\
    &\dot{g}_{\varphi\varphi}=\frac{d(r^2\sin^2(\theta))}{d\tau}=2r\dot{r}\sin^2(\theta)+r^2\sin(2\theta)\dot{\theta}, \\
    &\dot{u^{\theta}}=\frac{d^2u^{\theta}}{d\tau^2}=-\Gamma^{\theta}_{\mu\nu}u^{\mu}u^{\nu}+\frac{q}{m}F^{\theta\varphi}u_{\varphi}, \\
    &\dot{u^{\varphi}}=\frac{d^2u^{\varphi}}{d\tau^2}=-\Gamma^{\varphi}_{\mu\nu}u^{\mu}u^{\nu}+\frac{q}{m}F^{\varphi\theta}u_{\theta}.
    \end{split}
    \label{eq:prepare_absolute_angular_momentum_derivative}
\end{equation}

Expanding Eq.~\eqref{eq:absolute_angular_momentum_derivative} by Eq.~\eqref{eq:prepare_absolute_angular_momentum_derivative}, we get
\begin{equation}
\begin{split}
    \frac{dj}{d\tau}=&2r\dot{r}\dot{\theta}^2+2r^2\dot{\theta}\left ( -\Gamma^{\theta}_{\mu\nu}u^{\mu}u^{\nu}+\frac{q}{m}F^{\theta\varphi}r^2 \sin^2(\theta)\dot{\varphi} \right ) \\
& + dot{\varphi}^2\left[2r\dot{r}\sin^2(\theta)+2r^2\sin(\theta)\cos(\theta)\dot{\theta} \right] \\
&  + 2r^2\sin^2(\theta)\dot{\varphi}\left ( -\Gamma^{\varphi}_{\mu\nu}u^{\mu}u^{\nu}+\frac{q}{m}F^{\varphi\theta}r^2\dot{\theta} \right ).
\end{split}
\label{eq:expand_absolute_angular_momentum_derivative}
\end{equation}
Note that the 2 terms involving the electromagnetic tensor cancel out since $F^{\varphi\theta}=-F^{\theta\varphi}$. The  non-vanishing Christoffel symbols are 
\begin{equation}
\begin{split}
&\Gamma^{\varphi}_{\varphi r}=\Gamma^{\varphi}_{r\varphi}=1/r,\quad \Gamma^{\varphi}_{\theta \varphi}= \Gamma^{\varphi}_{\varphi\theta}=cot(\theta),\\
&\Gamma^{\theta}_{\varphi\varphi}=-\sin(\theta) \cos(\theta),\quad 
\Gamma^{\theta}_{\theta r}= \Gamma^{\theta}_{r\theta}=1/r.
\label{eq:static_chr_theta}
\end{split}
\end{equation}
Plugging them into Eq.~\eqref{eq:expand_absolute_angular_momentum_derivative}, we finally obtain
\begin{equation}
\begin{split}
    \frac{dj}{d\tau}=&r^2\dot{\theta}\left ( \sin(\theta)\cos(\theta)\dot{\varphi}^2-\frac{2}{r}\dot{\theta}\dot{\varphi} \right ) + 2r\dot{r}\sin^2(\theta)\dot{\varphi}^2 \\
& +2r\dot{r}\dot{\theta}^2+ r^2\sin(\theta)\cos(\theta)\dot{\theta}\dot{\varphi}^2 - 2r\dot{r}\sin^2(\theta)\dot{\varphi}^2 \\
& - 2r^2\sin(\theta)\cos(\theta)\dot{\theta}\dot{\varphi}^2=0.
\end{split}
\label{eq:result_absolute_angular_momentum_derivative}
\end{equation}
which proves the conservation of $j$.

Consequently, the radial equation reduces to\[
\dot{r}^2=\varepsilon^2-f(r)(1+\frac{j^2}{r^2}),
\]
with effective potential identical to Eq.~\eqref{eq:static_Veff}:
\begin{equation}
V_\text{eff}=f(r)(1+\frac{j^2}{r^2}).
\label{eq:static_effective_potential}
\end{equation}
The ISCO radius is then determined by the usual condition Eq.~\eqref{eq:static_ISCO}, identical to the neutral case. Thus, the particle charge does not shift the ISCO radius itself; rather, it controls the latitude of the orbit.

To solidify this connection, we can show that the extremum of the orbital latitude $\theta(r)$ occurs at the ISCO. The condition for an extremum is $d\theta/dr = 0$, which is equivalent to $d(\tan\theta)/dr = 0$. Applying this to Eq.~\eqref{eq:static_theta_r} gives
\begin{equation}
    \frac{d}{dr}\left(r\sqrt{\frac{\frac{M}{r}-\frac{P^2}{r^2}}{1-\frac{3M}{r}+\frac{2P^2}{r^2}}}\right) = 0.
\end{equation}
Evaluating the derivative yields the condition
\begin{equation}
    \frac{Mr^3 - 6M^2r^2 + 9MP^2r - 4P^4}{r^3} = 0.
\end{equation}
The numerator is exactly the polynomial from the ISCO condition in Eq.~\eqref{eq:static_ISCO}. Therefore, the maximum latitudinal deviation for a charged particle's orbit occurs precisely at the ISCO radius of a neutral particle.

\subsection{Synchrotron radiation of charged particles}\label{subsec:radiation}

Finally, we analyze the effect of synchrotron radiation on circular orbits, assessing the conditions under which these orbits can persist for sufficiently long lifetimes. In curved spacetime, the equation of motion of a charged particle incorporating the radiation-reaction effects reads~\cite{Bracken2024}
    \begin{align}
    m \ddot{z}^\alpha&=qF^\alpha_\beta \dot{z}^\beta + \frac{2}{3}q^2(\dddot{z}^\alpha - \ddot{z}^2 \dot{z}^\alpha) - \frac{1}{3}q^2(R^\alpha_\beta \dot{z}^\beta \nonumber\\&+ \dot{z}^\alpha R_{\beta\gamma}\dot{z}^\beta \dot{z}^\gamma) 
    + q^2 \dot{z}^\beta \int^\tau_{-\infty} \zeta^\alpha_{\beta\gamma} \ddot{z}^\gamma (\tau') d\tau' \nonumber\\
    & \equiv qF^\alpha_\beta \dot{z}^\beta + f^\alpha_\text{self}\,,
\label{eq:charged_self_force}
    \end{align}
where $z(\tau)$ is the particle trajectory, and overdots denote covariant derivatives with respect to proper time $\tau$ (giving the 4-velocity, 4-acceleration, etc.). The first term of the right-hand side is the usual Lorentz force, while $f^\alpha_\text{self}$ denotes the self-force for a generic curved background.

The timelike Killing vector $k^\mu = (1,0,0,0)$ introduced in Section~\ref{sec:static} allows us to define the energy of the particle as $E = -m\, k_\mu \dot{z}^\mu$. The emitted power is then found by computing
\begin{equation}
    \dot{E} = - k_\mu (m\ddot{z}^\mu) \, ,
\end{equation}
where we used the fact that $k^\mu$ satisfies the Killing equation $\nabla_\alpha k_\mu + \nabla_\mu k_\alpha = 0$.

Using Eq.~\eqref{eq:charged_self_force}, the total power of radiation is finally expressed as
\begin{equation}
    P_\text{tot}= -\dot{E} = k_\mu f^\mu_\text{self} = (f_\text{self})_0 \,.
\label{eq:radiation_gr_corrected}
\end{equation}
As expected, the Lorentz force yields no contribution to the emitted power.

Eq.~\eqref{eq:charged_self_force} involves the evaluation of the following non-local term,
\begin{equation}
  q^2 \dot{z}^\beta \int^\tau_{-\infty} \zeta_{0 \, \beta\gamma} \ddot{z}^\gamma (\tau') d\tau' \, ,
\label{eq:radiation_tail_term}
\end{equation}
arising from backscattering of radiation off the spacetime curvature. Here the quantity $\zeta$ is defined as
\begin{equation}
    \zeta_{\mu \nu \alpha}= v_{\mu \alpha; \nu} - v_{\nu \alpha; \mu},
    \label{eq:zeta_def}
\end{equation}
where $v_{\mu\nu}$ appears in the Hadamard form of the Green’s function,
\begin{equation}
    G^{(1)}_{\mu\nu}(x,z) = \frac{1}{4\pi^2} \left(\frac{u_{\mu\nu}}{\sigma} +v_{\mu\nu}\ln|\sigma| +w_{\mu\nu} \right)\, ,
\label{eq:Hadamard_Green_function}
\end{equation}
and the definitions of $\sigma$, $u_{\mu\nu}$ $w_{\mu\nu}$ can be found is Refs.~\cite{Najmi1985,DeWitt1960}. In the coincidence limit one has \cite{DeWitt1960}:
\begin{equation}
    \lim_{x \to z} v_{\mu\nu} = -\frac{1}{2} \delta^\alpha_\mu \left(R_{\alpha\nu} - \frac{1}{6}Rg_{\alpha\nu} \right).
    \label{eq:vtensor_approxima}
\end{equation}
The non-local term~\eqref{eq:radiation_tail_term} arises from the particle's own retarded field and is therefore a self-interaction effect. The local Larmor contribution, in contrast, is set by the particle's acceleration within the external MBH field. Since the test particle’s charge is presumably minuscule compared to the MBH’s magnetic charge, the contribution of the no-local term is only significant from the immediate past. We therefore evaluate Eq.~\eqref{eq:radiation_tail_term} by its coincident limit. 

For a charged particle moving on the circular orbit described in Section~\ref{sec:static}, the only non-vanishing acceleration component is 
\begin{equation}
    \ddot{z}^{\theta}=\pm \frac{Pq}{mr^3}\sqrt{\frac{\frac{M}{r}-\frac{P^2}{r^2}}{1-\frac{3M}{r}+\frac{2P^2}{r^2}}}.
    \label{eq:static_charged_acceleration}
\end{equation}
Where + stands for prograde case and - for retrograde case. Using the latter expression, one can verify that
\begin{equation}
    \lim_{x \to z} \zeta_{0 \, \beta\gamma} \ddot{z}^\gamma = 0 \, ,
\end{equation}
indicating that Eq.~\eqref{eq:radiation_tail_term} yields a negligible contribution.

Analogously, one can verify that $(\dddot{z})_0 = 0$ along circular obits. Therefore, the only non-vanishing terms in Eq.~\eqref{eq:radiation_gr_corrected} are
\begin{equation}
   P_\text{tot} =  -\frac{2}{3}q^2\ddot{z}^{\theta} \ddot{z}_{\theta} \dot{z}_0 -\frac{1}{3}q^2(R_{0\,\beta}\dot{z}^\beta+ \dot{z}_0 R_{\beta\gamma}\dot{z}^\beta \dot{z}^\gamma) \label{eq:static_charged_dot_acceleration} \,.
\end{equation}
We notice that in the limit of flat spacetime, $\dot{z}_0 \to -1$ and the curvature contribution vanishes, recovering the familiar Larmor formula. An explicit evaluation shows that 
\begin{align}
   P_\text{tot} &=  
\frac{
    2 P^2 q^2 \left( m^2 + q^2 \right) \left( \frac{M}{r} - \frac{P^2}{r^2} \right)
    }{3 m^2 r^4 \sqrt{ 1 -\frac{2M}{r} + \frac{P^2}{r^2} }} \nonumber \\
    & \times \left( \frac{1 -\frac{2M}{r} + \frac{P^2}{r^2} }{1 -\frac{3M}{r} + \frac{2P^2}{r^2} } \right)^{3/2}
\,.
\end{align}

We are interested in assessing the stability of orbits close to the ISCO, as this is relevant for black hole environments (such as accretion disks). The energy of the particle orbiting on the ISCO is $E= -m u_t(r_{\text{ISCO}})\sim \mathcal{O}(1) \times m$. Thus, the characteristic radiation timescale is estimated as
\begin{equation}
\delta\tau_{\text{rad}} \sim \frac{m}{P_\text{tot}(r_{\text{ISCO}})}\,. %\sim 1000(\frac{M}{M_\text{P}})^2 t_\text{P}.
\label{eq:radiation_time_test_particle}
\end{equation} 
For values $P/M \lesssim 0.5$ and $q/m \gg 1$, Eq.~\eqref{eq:radiation_time_test_particle} is approximated as
\begin{equation}
\delta\tau_{\text{rad}} \sim 3 \times 10^4 \, {\rm s} \left( \frac{M}{10^6 \, M_\odot}\right) \bigg(\frac{q}{m}\bigg)^{\!\!-4}
    \!\bigg(\frac{m}{M}\bigg)^{\!\!-1} \!\!\!\left(\frac{P}{M}\right)^{\!\!-2}
\end{equation}

\begin{figure}[htbp]
    \centering
\includegraphics[width=\linewidth]{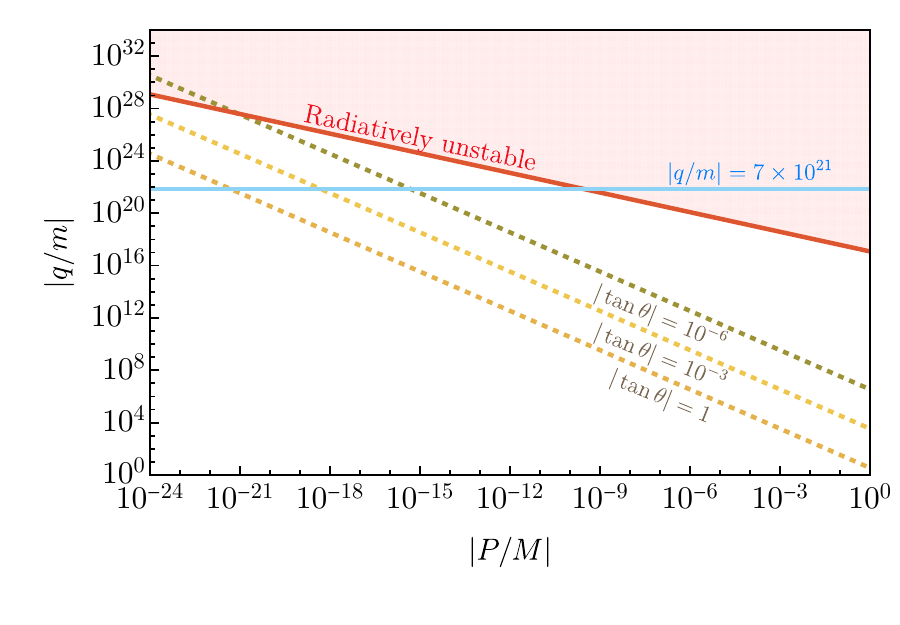}
    \caption{
    The parameter space of circular orbits for a charged particle with charge-to-mass ratio $|q/m|$ orbiting a massive black hole (MBH) with charge-to-mass ratio $|P/M|$. The dashed yellow lines indicate where the circular orbit exhibits significant latitudinal deviations ($|\tan\theta| = 1,10^{-3},10^{-6}$). The red region marks the values of the parameters for which the orbit becomes unstable under the emission of synchrotron radiation. The extent of this unstable region depends on the ratio $m/M$, as shown in Eq.~\eqref{eq:scalingratio}. In this figure, we take $m$ to be the electron mass and $M = 10^{6}\, M_\odot$.}
    \label{fig:valid_regime}
\end{figure}

For a circular orbit to be considered stable, its decay timescale must be much longer than its orbital period $t_\text{orb} = 2\pi / \omega$, where $\omega$ is defined in Eq.~\eqref{eq:static_angular_velocity}. Evaluating the ratio $\delta\tau_{\text{rad}} / t_\text{orb}$ at the ISCO, we find a simple scaling law valid for values $P/M \lesssim 10^{-2}$ and $q/m \gg 1$, i.e.
\begin{equation}\label{eq:scalingratio}
\frac{\delta\tau_{\text{rad}}}{t_\text{orb}} \sim 10^2 \,\bigg(\frac{q}{m}\bigg)^{\!\!-4}
    \bigg(\frac{m}{M}\bigg)^{\!\!-1} \!\!\left(\frac{P}{M}\right)^{\!\!-2}
\end{equation}

To visualize the parameter space where orbits are both stable and non-trivial, we plot the constraints in Figure~\ref{fig:valid_regime}. The dashed yellow lines correspond to constant values of $|\tan\theta|$, quantifying the latitudinal deviation caused by the magnetic field as per Eq.~\eqref{eq:scalingtheta}. Specifically, smaller values of $|\tan\theta|$ indicate orbits situated further from the equatorial plane. The red region delineates the area of parameter space where the timescale ratio $\delta\tau_{\text{rad}} / t_\text{orb} < 1$, marking the system as unstable to synchrotron radiation. This boundary is calculated for a benchmark system with $m$ equal to the electron mass and $M = 10^6 \, M_\odot$. Decreasing the mass ratio $m / M$ shifts the instability boundary (solid red line) upward in Figure~\ref{fig:valid_regime}, whereas the opposite trend occurs if we increase $m/M$.

% Section 3: theory of beta decay
\section{Orbital structure of Rotating Black Holes} \label{sec:rotate}

In this section, we extend our analysis to rotating black holes. The angular momentum of the black hole breaks spherical symmetry, reducing it to axial symmetry. Thus, we naturally align the $z$-axis with its rotation axis.

\subsection{Circular Orbits in Rotating MBHs}\label{subsec:rotate_MBHs}

The charged spinning spacetime is described by the Kerr–Newman metric \cite{Kerr1963, Mazur1982}, which is given by
\begin{equation}
\begin{split}
    &g_{\mu\nu}=\\
    &\begin{bmatrix}
 -\frac{\Delta-a^2\sin^2{\theta}}{\rho^2} & 0 & 0 & -\frac{a\sin^2{\theta}}{\rho^2}(2Mr-P^2)\\
 0 & \frac{\rho^2}{\Delta} & 0 & 0\\
 0 & 0 & \rho^2 & 0\\
 -\frac{a\sin^2{\theta}}{\rho^2}(2Mr-P^2) & 0 & 0 & \frac{(r^2+a^2)^2\sin^2{\theta}-\Delta a^2\sin^4{\theta}}{\rho^2}
\end{bmatrix},
\end{split}
\label{eq:rotate_metric}
\end{equation}
where
\begin{equation}
    \Delta=r^2-2Mr+a^2+P^2,\quad \rho^2=r^2+a^2\cos^2\theta,
    \label{eq:rotate_convenient_def}
\end{equation}
and $a=J/M$. The magnetic charge and angular momentum are constrained by
\begin{equation}
    P^2+a^2 \le M^2.
\label{eq:charge_rotate_constraint}
\end{equation}

The electromagnetic potential for the rotating MBH black holes is~\cite{Dyson2023}
\begin{equation}
    A_\mu= \frac{P\cos\theta}{\rho^2}\left(a, 0 , 0 ,- (r^2+a^2) \right) \, ,
    \label{eq:rotate_potential}
\end{equation}
which reproduces Eq.~\eqref{eq:non-rotating_potential} in the limit $a\to0$ up to a constant. We point the reader to Appendix~\ref{app:rotatingAmu} for a derivation of Eq.~\eqref{eq:rotate_potential}.

Following the same procedure outlined in Section~\ref{sec:static}, we solve the equations of motion under the assumption of a circular orbit,
$u^{\mu}=(u^t,0,0,u^\varphi)$,
where $u^\varphi$ is the constant angular velocity (its sign distinguishes between prograde and retrograde motion).

\begin{figure}[htbp]
    \centering
  \includegraphics[width=\linewidth]{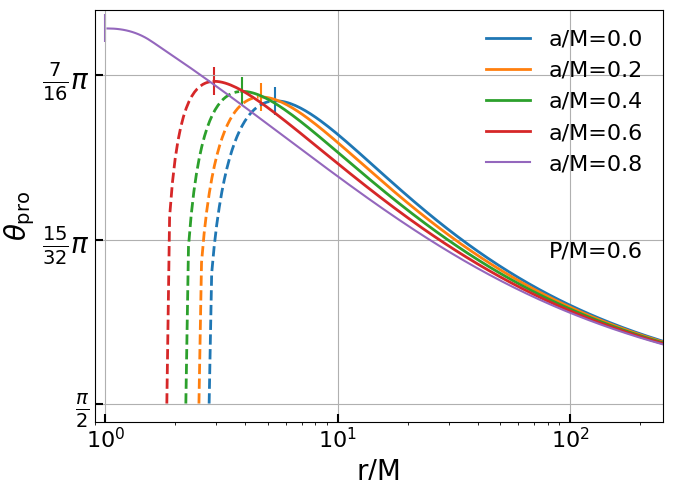} % File must exist!
  \caption{$\theta(r)$ of a $q/m=-1$ particle prograding a $P/M=0.6$ spinning MBH, for multiple $a/M$ values. Solid lines denote stable orbits, while dashed lines indicate instability against small perturbations; the transition point is marked on each curve. The y-axis is inverted so that north pole direction ($\theta = 0$) appears at the top of the figure, matching the intuitive orientation. The peak and smallest allowed radius decrease monotonously with $a/M$.}
  \label{fig:rotate_prograde_theta_r}
\end{figure}

\begin{figure}[htbp]
    \centering
  \includegraphics[width=\linewidth]{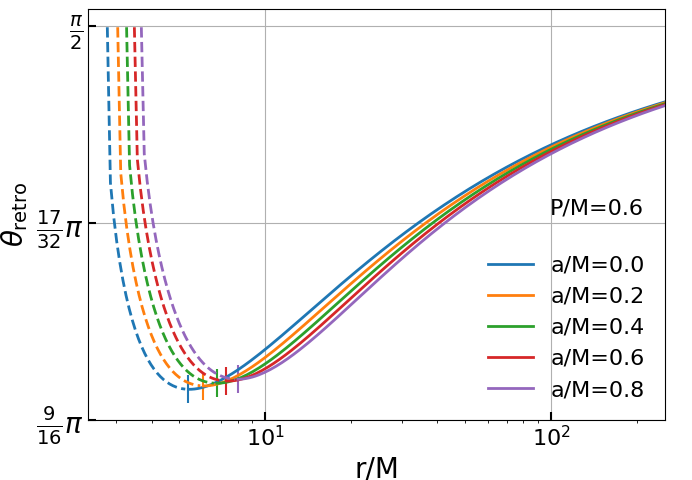} % File must exist!
  \caption{$\theta(r)$ of a $q/m=-1$ particle retrograding a $P/M=0.6$ spinning MBH, for multiple $a/M$ values. Solid lines denote stable orbits, while dashed lines indicate instability against small perturbations; the transition point is marked on each curve. The y-axis is inverted so that north pole direction ($\theta = 0$) appears at the top of the figure, matching the intuitive orientation. The peak and smallest allowed radius increase monotonously with $a/M$.}
  \label{fig:rotate_retrograde_theta_r}
\end{figure}

The $t$-component of the four velocity is determined imposing the constraint $g_{\mu\nu} u^\mu u^\nu = -1$, whereas Eq~\eqref{eq:static_eq_motion} leads to two independent equations that are used to numerically determine the polar angle $\theta$ and the radius $r$ of each orbit as a function of the parameter $u^\varphi$ under the condition $d^2 r / d\tau^2 = d^2 \theta / d\tau^2 = 0$ (see Appendix~\ref{app:eq_motion} for more details). For completeness, it is worth mentioning the Carter constant can be used to decouple the radial and polar equations \cite{Dyson2023} (see also Sec.~\ref{sec:ECBHs}). 
However, we proceed with numerically finding the roots of  Eq.~\eqref{eq:rotate_eq_motion_r} and Eq.~\eqref{eq:rotate_eq_motion_theta}, as this approach allows us to directly recover the family of circular orbits as a function of the spacetime coordinates $(r,\theta)$ and the parameters $(P/M, a/M, q/m)$ alone. The results for the prograde and retrograde branches are shown in Figure~\ref{fig:rotate_prograde_theta_r} and \ref{fig:rotate_retrograde_theta_r}, respectively. In each plot, we fix the particle charge-to-mass ratio at $q/m=-1$, set the magnetic charge to $P=0.6M$, and vary the rotation parameter from $a=0$ to $a=0.8M$, which corresponds to the extremal case according to Eq.~\eqref{eq:charge_rotate_constraint}. Analogous to the static case, we identify the transition from stable to unstable orbits and denote unstable regions with dashed lines. As will be discussed in Section~\ref{subsec:rotate_MBH_analysis}, this transition point still coincides with the peak radius of the orbital profile.

\begin{figure*}[htbp]
    \centering
  \includegraphics[width=1\linewidth]{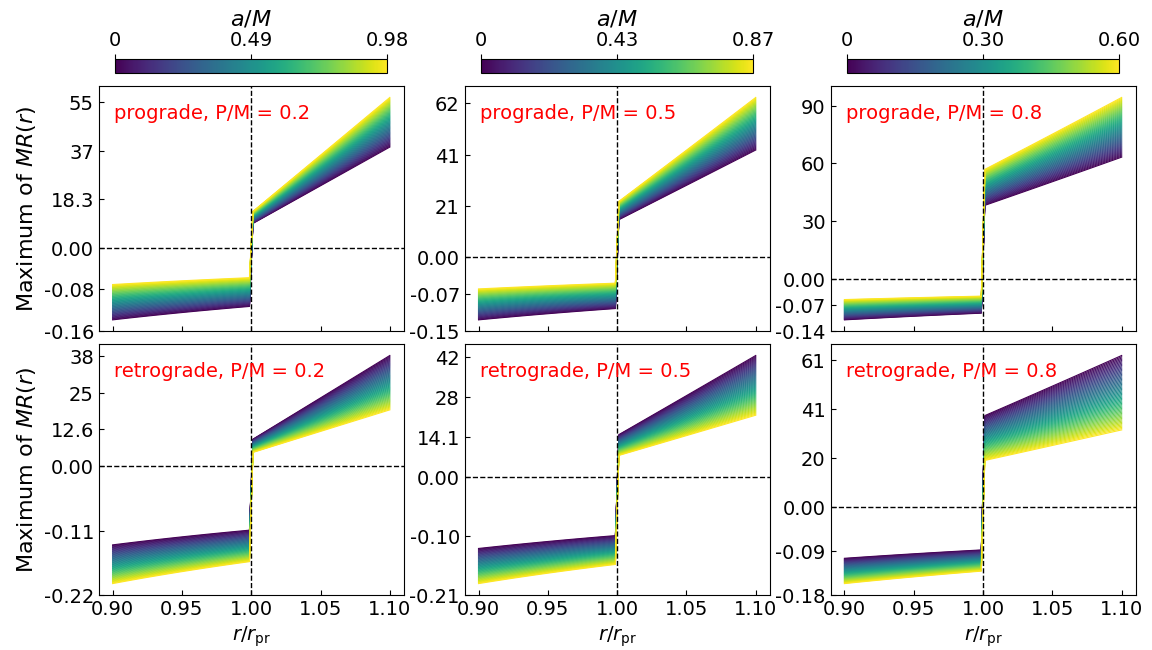} % File must exist!
  \caption{Maximum radial acceleration $MR(r,\theta,\omega)$ over all latitudes and $M\omega \leq 10$, as a function of the relative radius to photon ring radius $r/r_{\text{pr}}$. The black dashed lines mark $r=r_\text{pr}$ and $R=0$. A sharp transition occurs at the photon ring radius $r=r_{\text{pr}}$: $\max(R)<0$ for $r<r_{\text{pr}}$, prohibiting circular motion, while $\max(R)>0$ for $r>r_{\text{pr}}$, allowing circular orbits.}
  \label{fig:rotate_Rmax}
\end{figure*}

The behavior of $\theta(r)$ is qualitatively similar to the non-rotating case: it approaches $\pi/2$ in both the small- and large-radius limits, with a single intermediate peak. An exception arises for prograde motion around an extremal MBH, where $\theta$ attains its extremum near the smallest admissible radius. For retrograde orbits, both the peak and the innermost allowed radius occur farther from the black hole, consistent with the intuitive picture that prograde motion is more closely aligned with the spacetime geometry.

\subsection{Photon Rings and Stability Analysis}\label{subsec:rotate_MBH_analysis}

Following a similar argument as in Section~\ref{sec:static}, the photon ring radius in the equatorial plane is determined by the roots of the equation~\cite{Johannsen2013, Gal2019}
\begin{equation}
    r_{\text{pr}}^2 - 3Mr_{\text{pr}}+2P^2 \pm 2a\sqrt{Mr_{\text{pr}}-P^2} = 0,
    \label{eq:rotate_photon_ring_eq}
\end{equation}
where $+$ and $-$ denote prograde and retrograde motion, respectively. For extreme MBHs, $r_\text{pr}$ is identical to the horizon radius $r_\text{h}=M$. Our numerical exploration reveals that the photon ring continues to serve as the absolute inner boundary for circular orbits, in analogy with the non-rotating case. To demonstrate it, we analyze the radial component of the equation of motion, which determines the radial acceleration $R(r, \theta, \omega) = d^2r/d\tau^2$. A circular orbit can only exist if there is a set of parameters $(\theta, \omega)$ for which $R=0$. We numerically computed the maximum possible value of $R$ over all latitudes and a wide range of angular velocities at a given radius $r$. The results are presented in Figure~\ref{fig:rotate_Rmax}, where we show the maximum of $R(\theta, \omega)$ in the range of $0 \leq \theta \leq \pi,\quad0 \leq M\omega \leq 10$ as a function of the relative radius to photon ring radius ($r/r_\text{pr}$), for prograde and retrograde test particle with $q/m=-1$, for $a/M$ from $0$ to extremum (denoted by color from purple to yellow) at three different $P/M$. There is a jump point at $r_\text{pr}$: inside the photon ring radius, the maximum of $R(\theta, \omega)$ is negative, thus excluding any solution of a circular orbit; when $r>r_\text{pr}$ however, the maximum depends on the range of $\omega$ but tends to infinity as $\omega$ increases.

Consequently, $R(\theta, \omega)$ has no root at $r<r_\text{pr}$, and from Section~\ref{subsec:rotate_MBHs}, we see that Eq.~\eqref{eq:static_eq_motion} always have solution. Thus, the photon ring radius marks the critical radius above which there exist off-equatorial circular orbits for charged particles.

Unlike the static scenario, the classical Innermost Stable Circular Orbit (ISCO) is not uniquely defined for charged particles in rotating MBH spacetimes, as the radial motion depends on a third constant of motion (the Carter constant) in addition to energy and angular momentum. Consequently, a family of marginal stability radii exists for different Carter constants. We address this by defining a stability criterion specific to the class of orbits computed in Section~\ref{subsec:rotate_MBHs}.

We consider a charged test particle initially on an equilibrium circular orbit at $(r_0, \theta(r_0))$. We impose a small radial perturbation $r' = r_0 + \delta_r$ while constraining the particle to the manifold of circular solutions, such that the new polar angle is $\theta' = \theta(r')$. The perturbed angular velocity $\omega'$ is determined via angular momentum conservation. We then evaluate the radial acceleration at the perturbed radius $r'$ and define the stability coefficient $\kappa$ as the derivative of radial acceleration with respect to radial displacement
\begin{equation}
    \kappa(r_0) = \frac{d^2r/d\tau^2|_{r=r_0+\delta_r}}{\delta_r}.
    \label{eq:restore_force}
\end{equation}

\begin{figure}[htbp]
    \centering
  \includegraphics[width=\linewidth]{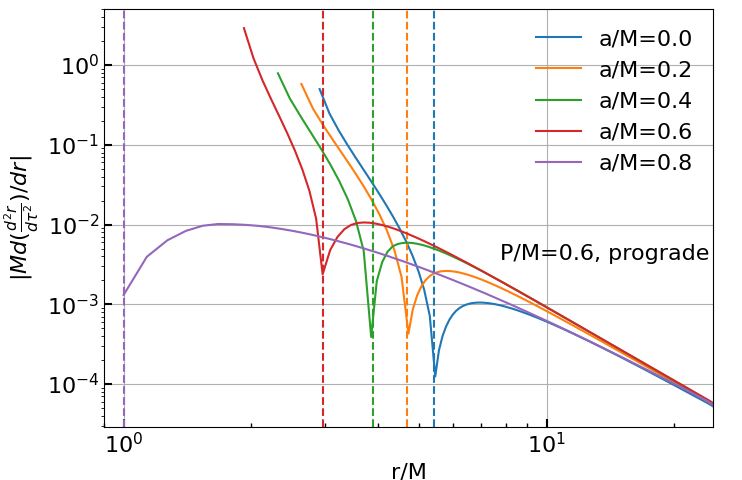}
  \caption{Normalized absolute value of stability coefficient $\kappa$ (log scale) versus radius for a prograde particle ($q/m=-1$) orbiting spinning MBHs ($P/M=0.6$). The function transitions from positive (unstable) to negative (stable), creating a sharp trough in the log plot. Dashed vertical lines indicate the peak radius of $\theta(r)$ for each spin parameter, which matches the stability transition point within numerical error.}
  \label{fig:rotate_prograde_stability}
\end{figure}

\begin{figure}[htbp]
    \centering
  \includegraphics[width=\linewidth]{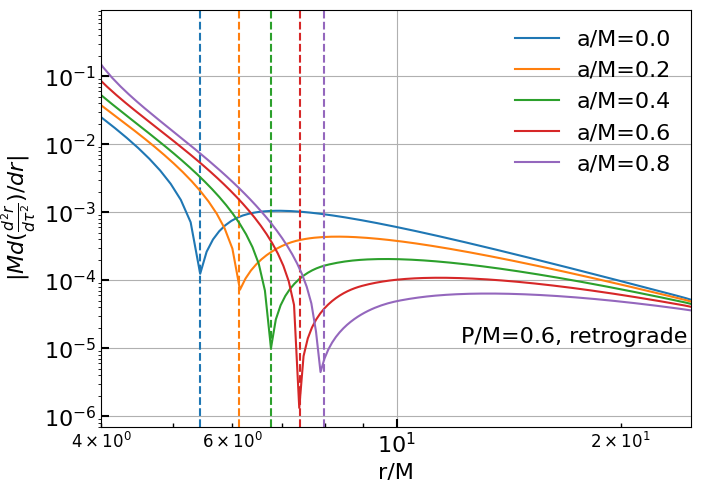}
  \caption{Normalized absolute value of stability coefficient $\kappa$ (log scale) versus radius for a retrograde particle ($q/m=-1$) orbiting spinning MBHs ($P/M=0.6$). Similar to the prograde case, the sharp trough corresponds to the sign change of the stability coefficient. The dashed vertical lines confirm that the stability transition coincides with the peak radius of the $\theta(r)$ profile.}
  \label{fig:rotate_retrograde_stability}
\end{figure}

A negative value of $\kappa$ indicates a restoring force, signaling stability against small perturbations, while a positive value implies instability. In Figures~\ref{fig:rotate_prograde_stability} and \ref{fig:rotate_retrograde_stability}, we plot the magnitude of $\kappa$ in log scale for the configuration used in Section~\ref{subsec:rotate_MBHs}. The force is consistently positive for radii smaller than the peak radius and negative for larger radii. The peak of the $\theta(r)$ profile therefore continues to demarcate the transition from unstable to stable orbits.

We therefore conclude that the minimum allowed radius for circular orbits coincides with the photon ring radius, and that the peak radius of the $\theta(r)$ function marks the innermost stable orbit.

\subsection{The case of electrically charged rotating black holes}\label{sec:ECBHs}

Finally, it is worth noting that rotating \textit{electrically} charged black holes also possess radial and polar components of both the electric and magnetic field. However, in contrast to the magnetically charged case, no off-equatorial circular orbits exist here. In other words, for an electrically charged particle, circular motion is confined to the equatorial plane only.

To see this, let us recall that the electromagnetic potential for the rotating electrically charged black hole is~\cite{Dyson2023}
\begin{equation}
    A_\mu= \frac{Q r}{\rho^2}\left(-1, 0 , 0 ,a \sin^2\theta \right) \, .
    \label{eq:rotate_potential_elec}
\end{equation}
While the energy $E$ and angular momentum $L$ per unit mass are conserved due to the existence of two Killing vectors, an additional constant of motion can be identified using the Hamilton–Jacobi approach. This quantity, known as the Carter constant~\cite{Carter:1968ks} and hereafter denoted as $\mathcal{C}$, ensures the integrability of the equations of motion. In particular, the equation for the polar angle can be written as~\cite{Dyson2023}
\begin{equation}
    \left(\frac{dz}{d\tau}\right)^2 \!\!\!= \!\frac{1}{\rho^4} \left\{\mathcal{C}(1 - z^2) - z^2 [L^2+ a^2 (1-z^2) (1-E^2)] \right\}\,,
\end{equation}
where $z = \cos \theta$. A circular orbit must satisfy $dz/d\tau = 0$ and $d^2 z / d\tau^2 = 0$. The former condition is used to express $\mathcal{C}$ in terms of the other parameters,
\begin{equation}\label{eq:Carter}
    \mathcal{C} = z^2 \left(-a^2
   \left(E^2-1\right)-\frac{L^2}{z^2-1}\right) \,. 
\end{equation}
Substituting Eq.~\eqref{eq:Carter} into the further condition $d^2 z / d\tau^2 = 0$, and asking that the orbit must be bound, i.e. $E^2 < 1$, leads to the unique solution $z = 0$. Thus, circular orbits in electrically charged rotating black holes exist only in the equatorial plane and are characterized by a vanishing Carter constant.

% Section 5: Current status of neutrino research
\section{Conclusions} \label{sec:conclusion}

In conclusion, we have presented a systematic and detailed analysis of the orbital dynamics of charged test particles around both static and rotating magnetically charged black holes. Although the motion of neutral particles in such spacetimes is well understood, we have demonstrated that the interplay between the gravitational field and the long-range magnetic Lorentz force gives rise to a family of stable, off-equatorial circular orbits. These trajectories, confined to a fixed latitude, represent a significant departure from the planar motion typically associated with black hole orbits.

Our primary results were derived for the static, spherically symmetric MBH. We obtained a closed-form analytical expression, Eq.~\eqref{eq:static_theta_r}, which determines the equilibrium polar angle $\theta$ of a charged particle as a function of its orbital radius. This relation establishes a direct link between fundamental radii: the minimum orbital radius corresponds to the photon ring, while the profile peak coincides with the ISCO.
Our analysis of synchrotron radiation indicates these orbits are viable over astrophysical timescales. We found that even electrons, which maximize radiative losses, can maintain a stable orbit with a significant latitudinal deviation around astrophysical-sized black holes.

We extended this investigation to rotating MBH spacetime, where we numerically computed the prograde and retrograde orbital branches. Our results show that the MBH spin modifies the orbits, while keeping the similar shape as the static case. Despite the added complexity and the loss of a direct ISCO connection, we demonstrated that the least allowed radius and the peak radius robustly maintain their roles as the photon ring and the inner boundary for stable circular motion, respectively. We also highlighted a crucial distinction by showing that such off-equatorial orbits are forbidden in electrically charged Kerr-Newman spacetimes, underscoring the unique role of the magnetic charge.

The existence of these stable, nonplanar orbits opens several avenues for future work. They may produce unique observational signatures, such as warps or gaps in accretion disks, or distinctive polarization patterns in the emitted synchrotron radiation. Future theoretical studies could explore the stability of these orbits against radial and vertical perturbations, include the effects of particle spin, or perform a more detailed analysis of the radiation reaction force. This work provides a clear and robust foundation for understanding the rich dynamics of charged matter in the vicinity of magnetically charged compact objects.

\section*{Data Availability}
The code used for calculations in this work is available at Ref.~\cite{CodeZenodo}.

% Acknowledgement
\section{Acknowledgments}

The authors gratefully acknowledge Julian Krolik and Emanuele Berti for their stimulating discussions. We thank David Pereñiguez for his insightful comments on the manuscript. L.X. would like to thank the Shanghai Jiao Tong University International Cooperation and Exchange Office and the Johns Hopkins University Global Education Office for facilitating a visit to JHU. L.D.G. is supported by NSF Grants No. PHY-2207502,
AST-2307146, PHY-090003 and PHY-20043, by NASA Grant No. 21-ATP21-0010, and by the John Templeton
Foundation Grant 62840. L.D.G. and D.K. are supported in part by the Simons Investigator Grant No. 144924, and D.K. is supported in part by the U.S.~National Science Foundation~(NSF) under Grant No.~PHY-1818899.

\appendix

\onecolumngrid
\section{Derivation of the electromagnetic tensor for the rotating MBH}
\label{app:rotatingAmu}
To derive the electromagnetic tensor in the rotating magnetically charged spacetime, we follow the complex transformation algorithm \cite{Newman1965, Adamo2014}, starting from the static spherically symmetric MBH described by the Reissner–Nordström metric. 

We first replace the time coordinate $t$ with the null coordinate $u$, defined through
\begin{equation}
    u=t-\int^r_0 \frac{dr'}{1-\frac{2M}{r'}+\frac{P^2}{r'^2}},
    \label{eq:def_null_time}
\end{equation}
such that the line element in Eq.~\eqref{eq:static_metric} simplifies to
\begin{equation}
    ds^2=-f(r)du^2-2\,du\,dr+r^2d\Omega^2.
    \label{eq:tetrad_line_element}
\end{equation}

Then we introduce the \textit{null tetrad} $\{l, n, m, \bar{m}\}$, where $l$ and $n$ are real while $m$ and $\bar{m}$ are complex conjugates. The only non-vanishing scalar products are $l \cdot n=1$ and $m \cdot \bar{m}=-1$. The contravariant metric can then be expressed in terms of the tetrad as
\begin{equation}
    g^{\mu\nu}=2 (m^{(\mu}\bar{m}^{\nu)} - l^{(\mu}n^{\nu)}).
    \label{eq:tetrad_metric}
\end{equation}
Comparing with Eq.~\eqref{eq:tetrad_line_element}, it is easy to write down the 1-forms, and hence tetrads of the null vectors:
\begin{equation}
\begin{split} 
 &l=\frac{\partial}{\partial r}, \\  
 &n=\frac{\partial}{\partial u} - \frac{1}{2}f(r)\frac{\partial}{\partial r},\\
 &m = \frac{1}{\sqrt{2} r}\frac{\partial}{\partial \theta}+\frac{i}{\sqrt{2} r\sin\theta}\frac{\partial}{\partial \varphi}.
 \label{eq:tetrad_one_form}
\end{split}
\end{equation}

\quad

The 4-potential is then expressed by the null vectors as
\begin{equation}
    A_\mu=\frac{-iP}{\sqrt{2}r\tan\theta}(m_\mu - \bar{m}_\mu).
    \label{eq:tetrad_static_potential}
\end{equation}

To obtain the spinning solution, we complexify the Reissner–Nordström tetrad:
\begin{equation}
\begin{split} 
 &l\to l'=\frac{\partial}{\partial r}, \\  
 &n\to n'=\frac{\partial}{\partial u} - \frac{1}{2}\left(1-\frac{M}{r}-\frac{M}{\bar{r}} + \frac{P^2}{|r|^2}\right)\frac{\partial}{\partial r},\\
 &m\to m' = \frac{1}{\sqrt{2} \bar{r}}\left(\frac{\partial}{\partial \theta}+\frac{i}{\sin\theta}\frac{\partial}{\partial \varphi}\right),
 \label{eq:tetrad_complexify}
\end{split}
\end{equation}
and simultaneously complexify the coordinates, analogously to the Schwarzschild-to-Kerr transformation:
\begin{equation}
\begin{split} 
 &r'=r + ia\cos\theta, \\  
 &u'=u - ia\cos\theta,\\
 &\theta' = \theta, \quad \varphi'=\varphi.
 \label{eq:coordinate_complexify}
\end{split}
\end{equation}

It is now straightforward to express the complex tetrad in the new coordinate system (Eq.~\eqref{eq:coordinate_complexify}) by an ordinary coordinate transformation (the primes indicating the new coordinates are dropped for simplicity):

\begin{equation}
\begin{split} 
 &l'=\frac{\partial}{\partial r},\quad n'=\frac{\partial}{\partial u} - \frac{1}{2}\left(1-\frac{2Mr - P^2}{\rho^2}\right)\frac{\partial}{\partial r},\\
 &m' = \frac{1}{\sqrt{2}(r + ia\cos\theta)} \\
 &\quad \quad \left( ia\sin\theta\frac{\partial}{\partial u} - ia\sin\theta\frac{\partial}{\partial r} + \frac{\partial}{\partial \theta} + \frac{i}{\sin\theta}\frac{\partial}{\partial \varphi} \right). 
 \label{eq:rotate_tetrad_complex}
\end{split}
\end{equation}

The electromagnetic 4-potential than follows from this null tetrad via Eq.~\eqref{eq:tetrad_metric} to be
\begin{equation}
    A_\mu=\left( \frac{Pa\cos\theta}{\rho^2} , 0 , 0 , -\frac{r^2+a^2}{\rho^2}P\cos\theta \right).
\end{equation}
which yields the covariant Maxwell tensor
%\begin{widetext}
\begin{equation}
    F_{\mu\nu}=\begin{bmatrix}
 0 & \frac{2Par \cdot \cos\theta}{\rho^4} & \frac{Pa (r^2-a^2\cos^2\theta) \sin\theta}{\rho^4} & 0\\
 -\frac{2Par \cdot \cos\theta}{\rho^4} & 0 & 0 & \frac{2Pa^2r\sin^2\theta \cos\theta}{\rho^4}\\
 -\frac{Pa (r^2-a^2\cos^2\theta) \sin\theta}{\rho^4} & 0 & 0 & \frac{P(r^2+a^2)(r^2-a^2\cos^2\theta)\sin\theta}{\rho^4}\\
 0 & -\frac{2Pa^2r\sin^2\theta \cos\theta}{\rho^4} & -\frac{P(r^2+a^2)(r^2-a^2\cos^2\theta)\sin\theta}{\rho^4} & 0
\end{bmatrix}.
    \label{eq:rotate_covariant_magnetic_tensor}
\end{equation}
%\end{widetext}
As in Section~\ref{sec:static}, we have defined the electromagnetic potential on the northern-hemisphere patch, while the southern patch is related by a gauge transformation. 

\onecolumngrid     % <--- switch to single column
\section{Equations of Motion of Charged Particles Around a Rotating MBH}
\label{app:eq_motion}
Consider a charged test particle moving on a circular orbit with 4-velocity
\[
u^\mu = (u^t, 0, 0, \omega).
\]
The normalization condition $u^\mu u_\mu = -1$ fixes $u^t$ as a function of $\omega$:
\begin{equation}
    u^t= \frac{-g_{t\varphi}\omega + \sqrt{(g_{t\varphi}\omega)^2 - g_{tt}g_{\varphi\varphi} \omega^2 - g_{tt}}}{g_{tt}},
    \label{eq:rotate_ut_omega_relation}
\end{equation}
where the metric coefficients are given in Eq.~\eqref{eq:rotate_metric}.

Expanding the radial component of the equation of motion with both the Christoffel symbols and the electromagnetic tensor Eq.~\eqref{eq:rotate_covariant_magnetic_tensor}, one finds the $r$-equation
\begin{equation}
    \begin{split}
    \frac{d^2r}{d\tau^2}= & \frac{\left(P^2+r (r-2 M)+1\right) \left(\cos (2 \theta ) M+M+2 r \left(P^2-M r\right)\right) (u^t)^2}{2 \left(r^2+\cos ^2\theta
   \right)^3}\\
   -& \frac{\left(P^2+r (r-2 M)+1\right) \omega  \left(M \cos ^2\theta +r \left(P^2-M r\right)\right) u^t\sin ^2\theta
   }{\left(r^2+\cos ^2\theta \right)^3}\\
   - & \frac{4 \left(P^2+r (r-2 M)+16\right) \omega  \left(16 M \cos ^2\theta +r \left(P^2-M
   r\right)\right) u^t\sin ^2\theta  }{\left(r^2+16 \cos ^2\theta \right)^3}\\
   + & \frac{\left(P^2+r (r-2 M)+16\right) \omega ^2 \left( \left(256 (M-r) \sin ^2\theta +r \left(r^4+16 \left(P^2-M r+16\right)
   \sin ^2\theta -256\right)+32 r \left(r^2+16\right)\right) \cos ^2\theta \right) }{\left(r^2+16 \cos ^2\theta \right)^3 \csc
   ^2\theta} \\
   + & \frac{q}{m \rho^2} \Biggl[a \left(2 M r-P^2\right) u^t
   \Biggl( \frac{2 a^2 P r \left(a^2+P^2+r^2-2 M r\right) \cos \theta  \left(\cos (2 \theta ) a^2+a^2+2 \left(P^2+r^2-2 M
   r\right)\right)}{\left(a^2+P^2+r (r-2 M)\right) \left(r^2+a^2 \cos ^2\theta \right)^3 \left(\cos (2 \theta ) a^2+a^2+2 r^2\right)} \\
   &\quad \quad \quad - \frac{4
   a^2 P r \left(P^2-2 M r\right) \left(a^2+P^2+r^2-2 M r\right) \cos \theta }{\left(a^2+P^2+r (r-2 M)\right) \left(r^2+a^2 \cos ^2\theta
   \right)^3 \left(\cos (2 \theta ) a^2+a^2+2 r^2\right)} \Biggr) \sin ^2\theta \\
   & \quad + a \left(2 M r-P^2\right)
   \omega \Biggl(\frac{4 P r \left(P^2-2 M r\right) \left(a^2+P^2+r^2-2 M r\right) \cos \theta  \sin ^2\theta  a^3}{\left(a^2+P^2+r (r-2
   M)\right) \left(r^2+a^2 \cos ^2\theta \right)^3 \left(\cos (2 \theta ) a^2+a^2+2 r^2\right)} \\
   &\quad \quad \quad +\frac{8 P r \left(a^2+P^2+r^2-2 M r\right)
   \cos \theta  \left(\left(a^2+r^2\right)^2-a^2 \left(a^2+P^2+r^2-2 M r\right) \sin ^2\theta \right) a}{\left(a^2+P^2+r (r-2 M)\right)
   \left(r^2+a^2 \cos ^2\theta \right)^2 \left(\cos (2 \theta ) a^2+a^2+2 r^2\right)^2}\Biggl) \sin ^2\theta \\
   & \quad  -\omega  \Biggl(\frac{2 a^2 P r \left(a^2+P^2+r^2-2 M r\right) \cos \theta  \left(\cos (2 \theta ) a^2+a^2+2 \left(P^2+r^2-2 M
   r\right)\right)}{\left(a^2+P^2+r (r-2 M)\right) \left(r^2+a^2 \cos ^2\theta \right)^3 \left(\cos (2 \theta ) a^2+a^2+2 r^2\right)} \\
   &\quad \quad \quad -\frac{4
   a^2 P r \left(P^2-2 M r\right) \left(a^2+P^2+r^2-2 M r\right) \cos \theta }{\left(a^2+P^2+r (r-2 M)\right) \left(r^2+a^2 \cos ^2\theta
   \right)^3 \left(\cos (2 \theta ) a^2+a^2+2 r^2\right)}\Biggl) \\
   & \quad \quad \quad \quad \times \left(\left(a^2+r^2\right)^2 \sin ^2\theta -a^2 \left(a^2+P^2+r^2-2 M
   r\right) \sin ^4\theta \right) \\
   & \quad +u^t \left(a^2\cos^2\theta+P^2+r^2-2 M r\right)
   \Biggl(\frac{4 P r \left(P^2-2 M r\right) \left(a^2+P^2+r^2-2 M r\right) \cos \theta  \sin ^2\theta  a^3}{\left(a^2+P^2+r (r-2 M)\right)
   \left(r^2+a^2 \cos ^2\theta \right)^3 \left(\cos (2 \theta ) a^2+a^2+2 r^2\right)} \\
   & \quad \quad \quad +\frac{8 P r \left(a^2+P^2+r^2-2 M r\right) \cos \theta
    \left(\left(a^2+r^2\right)^2-a^2 \left(a^2+P^2+r^2-2 M r\right) \sin ^2\theta \right) a}{\left(a^2+P^2+r (r-2 M)\right) \left(r^2+a^2
   \cos ^2\theta \right)^2 \left(\cos (2 \theta ) a^2+a^2+2 r^2\right)^2}\Biggl)\Biggl] = 0,
    \label{eq:rotate_eq_motion_r}
    \end{split}
\end{equation}

and the $\theta$-equation

\begin{equation}
    \begin{split}
    \frac{d^2\theta}{d\tau^2}= &\frac{\left(P^2-2 M r\right)
   \left(r^2+1\right) \omega  \cos \theta \sin \theta u^t }{\left(r^2+\cos ^2\theta\right)^3} -\frac{\left(P^2-2 M r\right) \cos \theta \sin \theta (u^t)^2}{\left(r^2+\cos ^2\theta\right)^3}+\frac{4 \left(P^2-2 M r\right)
   \left(r^2+16\right) \omega  \cos \theta \sin \theta u^t }{\left(r^2+16 \cos ^2\theta\right)^3} \\
   -&\frac{16 \omega ^2 \cos \theta
   \sin ^3\theta \left(16 \left(P^2+r^2-2 M r+16\right) \sin ^2\theta-\left(r^2+16\right)^2\right)}{ \left(r^2+16 \cos ^2\theta \right)^3} \\
   +&\frac{\omega ^2 \left(r^2+16 \cos ^2\theta\right)
   \left(\left(r^2+16\right)^2-32 \left(P^2+r^2-2 M r+16\right) \sin ^2\theta\right) \sin (2 \theta )}{2 \left(r^2+16 \cos ^2\theta \right)^3} \\
   +&\frac{q}{m\rho^2} \Biggl[a \left(2
   M r-P^2\right) u^t  \Biggl(\frac{P \left(a^2+r^2\right) \left(r^2-a^2 \cos ^2\theta\right) \left(\cos (2 \theta ) a^2+a^2+2
   \left(P^2+r^2-2 M r\right)\right) \csc \theta}{\left(a^2+P^2+r (r-2 M)\right) \left(r^2+a^2 \cos ^2\theta\right)^3 \left(\cos (2
   \theta ) a^2+a^2+2 r^2\right)}\\
   &\quad\quad\quad-\frac{2 a^2 P \left(P^2-2 M r\right) \left(r^2-a^2 \cos ^2\theta\right) \sin \theta}{\left(a^2+P^2+r
   (r-2 M)\right) \left(r^2+a^2 \cos ^2\theta\right)^3 \left(\cos (2 \theta ) a^2+a^2+2 r^2\right)}\Biggl) \sin ^2\theta\\
   &\quad+a \left(2 M r-P^2\right) \omega  \Biggl(\frac{4 a P \left(r^2-a^2 \cos ^2\theta\right) \left(\left(a^2+r^2\right)^2-a^2
   \left(a^2+P^2+r^2-2 M r\right) \sin ^2\theta\right) \sin \theta}{\left(a^2+P^2+r (r-2 M)\right) \left(r^2+a^2 \cos ^2\theta
   \right)^2 \left(\cos (2 \theta ) a^2+a^2+2 r^2\right)^2}\\
   &\quad\quad\quad+\frac{2 a P \left(P^2-2 M r\right) \left(a^2+r^2\right) \left(r^2-a^2 \cos
   ^2\theta\right) \sin \theta}{\left(a^2+P^2+r (r-2 M)\right) \left(r^2+a^2 \cos ^2\theta\right)^3 \left(\cos (2 \theta ) a^2+a^2+2
   r^2\right)}\Biggl) \sin ^2\theta\\
   &\quad-\omega \left(\left(a^2+r^2\right)^2 \sin ^2\theta-a^2 \left(a^2+P^2+r^2-2 M r\right) \sin ^4\theta\right)\\
   &\quad\quad\times \Biggl(\frac{P \left(a^2+r^2\right) \left(r^2-a^2 \cos ^2\theta
   \right) \left(\cos (2 \theta ) a^2+a^2+2 \left(P^2+r^2-2 M r\right)\right) \csc \theta}{\left(a^2+P^2+r (r-2 M)\right) \left(r^2+a^2
   \cos ^2\theta\right)^3 \left(\cos (2 \theta ) a^2+a^2+2 r^2\right)}\\
   &\quad\quad\quad-\frac{2 a^2 P \left(P^2-2 M r\right) \left(r^2-a^2 \cos ^2\theta
   \right) \sin \theta}{\left(a^2+P^2+r (r-2 M)\right) \left(r^2+a^2 \cos ^2\theta\right)^3 \left(\cos (2 \theta ) a^2+a^2+2
   r^2\right)}\Biggl) \\
   &\quad+u^t  \left(a^2\cos^2\theta +P^2+r^2-2 M r\right) \Biggl(\frac{4 a P \left(r^2-a^2 \cos ^2\theta\right)
   \left(\left(a^2+r^2\right)^2-a^2 \left(a^2+P^2+r^2-2 M r\right) \sin ^2\theta\right) \sin \theta}{\left(a^2+P^2+r (r-2 M)\right)
   \left(r^2+a^2 \cos ^2\theta\right)^2 \left(\cos (2 \theta ) a^2+a^2+2 r^2\right)^2}\\
   &\quad\quad\quad+\frac{2 a P \left(P^2-2 M r\right)
   \left(a^2+r^2\right) \left(r^2-a^2 \cos ^2\theta\right) \sin \theta}{\left(a^2+P^2+r (r-2 M)\right) \left(r^2+a^2 \cos ^2\theta
   \right)^3 \left(\cos (2 \theta ) a^2+a^2+2 r^2\right)}\Biggl) \Biggl]= 0.
    \label{eq:rotate_eq_motion_theta}
    \end{split}
\end{equation}

Where $\rho$ is defined in Eq.~\eqref{eq:rotate_convenient_def} and the explicit form of $u^t$ is presented above in Eq.~\eqref{eq:rotate_ut_omega_relation}.
Together, the $r$- and $\theta$-equations determine admissible pairs $(\theta,\omega)$ at each orbital radius. For every radius larger than the photon-ring radius, the equations yield two physical solutions for $\omega$: one positive (prograde motion) and one negative (retrograde motion).  

In Kerr-Newman spacetime the equations take a similar form. However, all circular solutions are confined to $\theta = \pi/2$.

\section{More on electrically charged rotating black holes}

\begin{figure*}[htbp]
    \centering
    \includegraphics[width=1\linewidth]{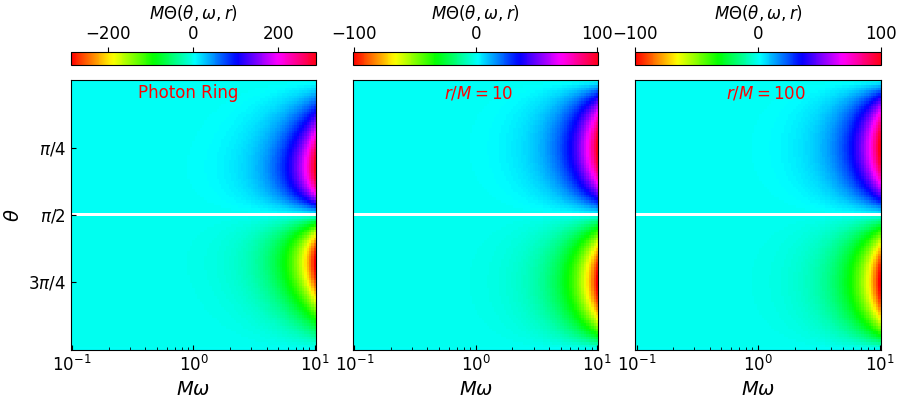} % File must exist!
    \caption{$\Theta(\theta, \omega, r)$ in Kerr–Newman space-time with $a/M = Q/M = 0.5$, evaluated at a wide range of $\omega$, at three radii: the photon ring ($r=r_\text{pr}=2.12M$), $r/M=10$, and $r/M=100$, for a prograde $q/m=-1$ particle. Regions where $\Theta=0$ are masked in white.
    At all radii, $\Theta$ vanishes only in the equatorial limit $\theta \to \pi/2$, thereby precluding the existence of off-equatorial circular orbits.}
\label{fig:electric_rotate_Theta_nozero}
\end{figure*}
The electromagnetic tensor in Kerr–Newman spacetime takes the form
\begin{widetext}
\begin{equation}
    F^{(KN)}_{\mu\nu}=\begin{bmatrix}
 0 & -\frac{Q(r^2 - a^2 \cos^2\theta)}{\rho^4} & \frac{2Qra^2\sin\theta \cos\theta}{\rho^4} & 0\\
 \frac{Q(r^2 - a^2 \cos^2\theta)}{\rho^4} & 0 & 0 & -\frac{Qa\sin^2\theta (r^2 - a^2 \cos^2\theta)}{\rho^4}\\
 -\frac{2Qra^2\sin\theta \cos\theta}{\rho^4} & 0 & 0 & \frac{2Qra(r^2+a^2)\sin\theta \cos\theta}{\rho^4}\\
 0 & \frac{Qa\sin^2\theta (r^2 - a^2 \cos^2\theta)}{\rho^4} & \frac{2Qra(r^2+a^2)\sin\theta \cos\theta}{\rho^4} & 0
\end{bmatrix}.
    \label{eq:rotate_covariant_electric_tensor}
\end{equation}
\end{widetext}
Both $F^{(KN)}_{t\theta}$ and $F^{(KN)}_{\theta\varphi}$ vanish at $\theta = \pi/2$, since they contain an overall factor of $\cos\theta$. Thus, on the equatorial plane, only a radial electric field and a polar magnetic field remain, yielding a purely radial electromagnetic force that admits equatorial orbits. When $\theta \neq \pi/2$, however, the $\theta$-component of Eq.~\eqref{eq:static_eq_motion} has no physical solution for $\theta$ and $\omega$, thereby excluding off-equatorial circular orbits. 

To show this explicitly, we define 
\begin{equation}
\Theta(\theta, \omega, r) = d^2\theta/d\tau^2\,,
\end{equation}
as the total polar acceleration. Figure~\ref{fig:electric_rotate_Theta_nozero} plots $\Theta$ for a charged particle in an electrically charged KN spacetime. Away from the equatorial plane, $\Theta$ is strictly non-zero, confirming that no equilibrium solution exists. Furthermore, the sign of $\Theta$ indicates a restoring force ($\Theta>0$ for $\theta<\pi/2$ and $\Theta<0$ for $\theta>\pi/2$), that drives the particle back toward the equatorial plane($\theta=\pi/2$). This result highlights a fundamental difference: the purely radial field of a magnetic monopole provides a unique mechanism for latitudinal confinement that is absent in the field of a rotating electric charge.

% Reference section

\bibliographystyle{apsrev4-2}

\bibliography{sections/references.bib}

\end{document}